\documentclass{article}
\usepackage{ijcai22}

\usepackage{times}
\usepackage{soul}
\usepackage{url}

\usepackage[utf8]{inputenc}
\usepackage[small]{caption}

\usepackage{amsmath}
\usepackage{amsthm}
\usepackage{booktabs}
\usepackage{microtype}
\pdfoutput=1
\usepackage{graphicx}

\title{Cooperative Marine Operations via Ad Hoc Teams}

\author{
Ignacio Carlucho$^1$\and
Arrasy Rahman$^1$\and
William Ard$^2$\and
Elliot Fosong$^1$\and\newline
Corina Barbalata$^2$ \And
Stefano V. Albrecht$^1$
\affiliations
$^1$School of Informatics, University of Edinburgh, UK \\
$^2$Department of Mechanical \& Industrial Engineering, Louisiana State University, Baton Rouge, USA
\emails
\{ignacio.carlucho, arrasy.rahman, e.fosong, s.albrecht\}@ed.ac.uk \\
\{ward2, cbarbalata\}@lsu.edu
}

\begin{document}

\maketitle

\begin{abstract}

While research in ad hoc teamwork has great potential for solving real-world robotic applications, most developments so far have been focusing on environments with simple dynamics. In this article, we discuss how the problem of ad hoc teamwork can be of special interest for marine robotics and how it can aid marine operations. Particularly, we present a set of challenges that need to be addressed for achieving ad hoc teamwork in underwater environments and we discuss possible solutions based on current state-of-the-art developments in the ad hoc teamwork literature.

\end{abstract}

\section{Introduction}

Marine environments offer a number of valuable resources, and while oceans cover a large percentage of the earth surface, most of the underwater environments remain unexplored. 
In recent years, governmental efforts have been carried out to increase human presence in the oceans, in order to explore and research underwater environments as well as to exploit natural resources \cite{egan2021exploration,amon2022my}. However the high cost of operation in underwater environments still presents a great barrier for large scale operations. Typical monitoring and surveying systems utilize remotely operated vehicles (ROVs), but those require an operator which controls the system from a ship, however such systems reportedly
costs in the order of £10K per day \cite{TEAGUE2018333}.

To lower operational cost, researchers are turning to autonomous underwater vehicles (AUVs) and autonomous surface vehicles (ASVs), which do not require a human operator to complete tasks.
However, the sensing capabilities of AUVs are constrained by their often smaller size and battery capacity, which limits the possible payload. 
Furthermore, due to the lack of global positioning system in underwater environments, AUVs are prone to drift and limited positioning accuracy \cite{Maurelli2022}.
Researchers address this precision problem by effectively increasing accuracy of agents through the deployment of teams of robots that share navigational information between team members, which takes advantage of the low cost of these system \cite{Jonatan2019}. 
At this same time, this has opened possibilities for new types of collaborative missions \cite{Towardsmixedinitiatives2011}. For instance, a team of AUVs can be used for survey missions, where their combined perception could increase the accuracy of tasks such as seabed mapping or underwater exploration.

Most of multi-agent research so far has focused on cases in which cooperation pertains to a set of known teammates which have been trained together, such as in multi-agent reinforcement learning (MARL) \cite{Papoudakis2019DealingWN}. However, underwater operations, especially rescue missions and disaster recovery missions, require the need of cooperation without prior coordination, as was the case for the search and rescue operations carried out for the ARA San Juan\footnote{\url{https://en.wikipedia.org/wiki/ARA_San_Juan_(S-42)}} \cite{de2019hundimiento}. 
The ARA San Juan was a submarine of the Argentinian navy that suddenly lost communication and became lost in the sea. A search and rescue task started immediately to locate the submarine and rescue the crew, however the search area was extremely large and multiple sensors or sensing units were needed. The international community cooperated, by providing many different AUVs, and other vessels to try to locate the lost submarine \cite{zhongming2017whoi,Nielsen2021}. However, many of these systems were from different manufacturers, operating independently, which made search efforts chaotic. Multiple areas where surveyed repeatedly, information was not shared between vehicles, all of which made the search inefficient.

We argue here that the development of autonomous marine vehicles capable of ad hoc teamwork \cite{mirskySurveyAdHoc2022} will be useful for marine operations. 
Particularly, we believe that search and rescue operations, as well as intervention operations, could be greatly improved by agents who are capable of cooperating on the fly with unknown teammates with different payloads and characteristics. 
However, to achieve such level of autonomy, a set of challenges need to be addressed. In the following sections, we will introduce current developments in the field of ad hoc teamwork, and  we will discuss what we believe are the most difficult challenges to be faced. Finally, we will provide a discussion regarding future steps necessary to solve the challenge.

\section{Related Works}

Control strategies for multi-agent systems have been widely studied in the literature \cite{Gronauer2022}.
In particular, for marine environments, several strategies for coordination and formation control have been developed \cite{Learntonavigate2019}. This problem has been also studied in vehicles that are subject to communication delays \cite{formationcontrol}.
Recent works have also demonstrated the advantages of utilizing a network of underwater robots for conducting missions. In \cite{Rahmati_2019}, MARL was used for developing a network of robots that could monitor the quality of water in real time. However, MARL approaches typically assume that teammates are trained together and thus information about teammates is available.

Ad hoc teamwork is the problem of developing agents that can cooperate on the  fly with new teammates \cite{Stoneadhoc,mirskySurveyAdHoc2022}. Ad hoc learners  typically use type-based methods that associate teammates to a certain type \cite{ALBRECHT201866}, and then use action selection mechanisms to select the best action \cite{EHBA}. 
More recently, researchers have also leveraged graph neural networks allowing them to work with an unknown number of teammates \cite{GPL}. Other avenues of research have studied how communication can improve collaboration in teams of unknown agents \cite{mirskySurveyAdHoc2022}.

\section{Challenges}

We are interested in developing an agent, usually called the learner, which is able to cooperate on the fly with an unknown number of teammates for solving mission in marine operations. The learner will face several challenges:
i) Perception capabilities might vary between vehicles, some vehicles might have sonars, cameras, and/or lidars. 
ii) Communication in underwater environments have delays and low bandwidth, which limits the data transfer capabilities of these agents. 
iii) Each agent can only observe part of the environment, and most of the sensor readings are noisy. This also includes accurate self localization issues for most of the underwater vehicles. 
iv) Agents might differ in their mobility and intervention capabilities. For instance, AUVs can be underwater and move in 4, or sometimes 6 degrees of freedom, while ASVs can only move on the surface. Others might not even be able to move, such as buoys. 
iv) In ad hoc search and rescue operations, the number of teammates will be unknown. This makes the typical approaches not directly applicable.
vi) Agents should be able to work autonomously and cooperate on the fly to solve the new task without relying on previously shared tasks and communication protocols.

In the next section, we will discuss in more detail why these challenges present such difficulty for marine robotics, and examine possible approaches that would enable the development of real-world ad hoc agents.

\section{Future Developments}

\looseness=-1
The perceptual capabilities of robots in underwater environments are extremely limited. Therefore, agents should make use of communication channels in order to probe other agents and decrease the uncertainty of the current state \cite{ALA16-grizou}.
While communicating agents in ad hoc teamwork have been presented before \cite{Apenny}, this will pose a much harder challenge in underwater environments, due to issues related to communication delays and low bandwidth \cite{Bouk2016}.  
We desire algorithms that enable agents to deal with communication uncertainty, such as those caused by attenuation or delays \cite{RAHMAN202179}, and that at the same time can improve cooperation in ad hoc settings.

Additionally, underwater environments require complex control systems that can deal with the highly nonlinear dynamics, uncertainty, and variable working conditions of the marine environments \cite{CARLUCHO2021102726}. This contrasts with ad hoc teamwork methodologies that have been mostly developed for agents with limited number of actions and simple dynamics. One way to integrate these differences will be to generate a control hierarchy, consisting of a set of layers, such as in \cite{motorcontrol}. In this way, the higher layers of the hierarchy are meant to solve the decision making and coordination problem by means of advanced reinforcement learning (RL) algorithms \cite{GPL}, while low-level dynamics are handled by classical control modules or by hybrid modules that integrate both classical control and artificial intelligence techniques \cite{CARLUCHO2020280}. 

As previously stated, agents should be able to cooperate with an unknown number of teammates of different types, as is the case for applications such as the one described in the motivating example in the introduction. Recent methods have shown how graph neural networks could be leveraged to develop coordination in such cases \cite{GPL}, however have done so having full observability of the state. Methods that can robustly handle partial and noisy observations are yet to be developed.

While control policies can allow agents to coordinate and carry out intervention procedures, such as rescue operations, we envision missions in which agents need to combine their perceptions capabilities to survey, monitor, or search for a specific object of interest. 
These missions will require teams of hybrid robots with different sensing and motor capabilities, and possibly even under a mixture of human-robot teams. In these cases, robots will be required to recognize their own skills and those of their team member in order to exploit each individual strength \cite{Bosch2019SixCF}. 
But more importantly, robots in the teams should also be able to co-create knowledge \cite{Hautala2022}. 

In addition to the previously described problems we also anticipate the need for the development of realistic simulation environments.  
Simulators should be able to model hydrodynamic effects, communication channels including delays and realistic sensors, such as cameras and sonars. 
Additionally, it should be possible to simulate different platforms, such as AUVs and underwater manipulators, as well as ASVs. These environments also should allow to model different types of tasks such as interventions or underwater surveying.

\section{Conclusion}
In this work we discussed how ad hoc teamwork can aid in the development of marine operations, specifically in search and rescue scenarios. We identified the major challenges in the development of real world ad hoc agents for underwater operations, and we discussed future developments that are necessary for achieving these types of autonomous agents.

\bibliographystyle{ieeetr}
\bibliography{main}

\end{document}